        \documentstyle[11pt]{article}
\def\lsim{\, \lower2truept\hbox{${<
\atop\hbox{\raise4truept\hbox{$\sim$}}}$}\,}
\def\gsim{\, \lower2truept\hbox{${>
\atop\hbox{\raise4truept\hbox{$\sim$}}}$}\,}
\def\frac#1/#2{\leavevmode\kern.1em
\raise.5ex\hbox{\the\scriptfont0 #1}\kern-.1em
/\kern-.15em\lower.25ex\hbox{\the\scriptfont0 #2}}

\begin{document}

\begin{center}

{\Large Background radiation}

\end{center}

\bigskip

\begin{center}

G. De Zotti, A. Franceschini, L. Toffolatti, P. Mazzei

Osservatorio Astronomico di Padova

L. Danese

Dipartimento di Astronomia, Padova

\end{center}

\bigskip\medskip
\section{Introduction}

We present a short overview on the extragalactic
background radiation from radio to X-rays,
with an eye to the relation to galaxy formation. The emphasis is on
astrophysical (as opposed to cosmological) backgrounds. In particular,
we will only cursorily deal with the cosmic microwave background (CMB)
and with reference only to spectral distortions and anisotropies due to
gas processes at the formation of collapsed structures.

\section{The radio background}

The extragalactic radio background has received little attention in the last
25 years, while the interest of radio astronomers was shifted on one side
towards accurate measurements of the CMB and, on the
other side, towards deeper and deeper surveys. In fact, it may well be
that most of the extragalactic radio background has already been resolved
into sources. Nevertheless, improved determinations of the intensity of
the radio background are still of considerable interest, for at least
three reasons: i)
a significant, or even large, fraction of it may not come from discrete
sources at all but from emission of diffuse intergalactic or pre-galactic
material; ii)
even if all the background is actually attributable to discrete sources,
an indefinite fraction of them may be too faint to be detected even in the
deepest surveys;
ii) the integrated emission of extragalactic sources and from a pre-- or
inter--galactic medium constitutes a ``foreground'' signal which needs to
be determined and subtracted out to infer the CMB spectrum.

Unfortunately, its intensity is difficult to measure because it is swamped
by the non thermal
radiation of our own Galaxy and/or by the CMB (the minimum sky brigthness
at 178 MHz is $\simeq 80\,$K, while that of the isotropic component
is estimated to be in the range 15--37 K: Longair \& Sunyaev 1971).
However, improving such estimates is possible even with the
classic ``T-T plot'' method.

The wealth of data on source counts and related statistics at many frequencies
(notably at 1.4 and 5 GHz where counts have been determined down to
$\sim 10\mu$Jy) makes possible direct estimates of the
contributions  from extragalactic
sources. At 178 MHz the  result is $\simeq
22\,$K; $\simeq 60\%$ of the flux comes from radiogalaxies and
quasars, the remaining part from sub-mJy radio sources, a large fraction of
which is likely to be made up by active star-forming galaxies.

At high frequencies, the radio background is dominated by compact sources,
which are generally QSO's. Preliminary identifications of
sources detected by the EGRET instrument on the Compton Observatory
seem to indicate that compact radio quasars also dominate the
extragalactic $\gamma$-ray sky (Bignami 1992). Their
$\gamma$-ray to radio flux ratios  are generally
well above the corresponding ratio of background intensities
[$(\nu I_\nu)_{{\rm bkg,}\, 100\,{\rm MeV}}/(\nu I_\nu)_{{\rm bkg,}\, 10\,
{\rm GHz}}\simeq 5$] suggesting that this kind
of sources may
be the dominant contributors to the $\gamma$-ray background.

\section{The far-IR/sub--mm background}

\subsection{Comptonization distortions of the CMB}

A significant sub-mm excess can be produced by comptonization of microwave
photons off hot electrons.
An order of magnitude estimate of CMB spectral distortions
expected in different scenarios can be easily obtained by scaling the
well known result that an intergalactic medium (IGM)
having a kinetic energy density, referred
to the present time, $\epsilon_{\rm IGM}(z=0) \approx 10^{-13}\,
\hbox{erg}\,\hbox{cm}^{-3}$ ($H_0 = 50$), sufficient to produce the XRB, yields
$y = \int(kT_e/mc^2)n_e \sigma_Tc\,dt \approx 10^{-2}$ (Field \& Perrenod
1977).
Well known energy sources include:

\begin{itemize}

\item {\it Stellar nucleosynthesis}. The present average density of metals
in observed galaxies is estimated to be (Songaila et al. 1990):
$6\times 10^{-5} < (H_0/50)^2 \Omega < 2\times 10^{-4}$. If the
average density of helium synthesized in stars is $\Omega_{He} \simeq
3\Omega_Z$, the energy density produced by stellar nucleosynthesis is
$\epsilon_\star(z=0) \approx 2\hbox{--}6 \times 10^{-15}\,
\hbox{erg}\,\hbox{cm}^{-3}$ ($z_\star \simeq 2$). Only a fraction of this
contributes to heating of the IGM.

\item {\it Binding energy}. The binding energy of baryons in galaxies is
$\epsilon_{b}(z=0) \simeq \frac1/2 \rho_b v^2(1+z_{\rm coll})^{-1}
\approx 2\times 10^{-18} [10/(1+z_{\rm coll})]\,
\hbox{erg}\,\hbox{cm}^{-3}$. The energy density associated with larger
scale structure is probably of the same order.

\item {\it Nuclear activity}. The
energy density produced by AGNs, estimated from B counts assuming
a ratio of bolometric to B flux $\kappa = 30$
and an effective redshift $z_{\rm AGN} = 1.5$ is
(Padovani et al. 1990): ${\epsilon_{\rm AGN}(z=0) \approx 5\times 10^{-16}\,
\hbox{erg}\,\hbox{cm}^{-3}}$,
about \frac1/6 of which is in the form of ionizing photons.

\end{itemize}

Although all the above values are highly uncertain, it appears likely,
also in view of the primordial nucleosynthesis constraint ($\Omega_b
\lsim 0.1$), that the amount of energy that can have been released
by well established energy sources to heat up the IGM is bounded by
$\epsilon_{\rm IGM}(z=0) \lsim
{\rm few} \times 10^{-15}\,\hbox{erg}\,\hbox{cm}^{-3}$, so that
$y$ is expected to be smaller (and possibly much smaller)
than  ${\rm few} \times 10^{-4}$.

Detailed calculations of comptonization distortions from gas heated
during the formation of large scale structures have been carried out
by Cen \& Ostriker (1992), Cavaliere et al. (1991), Markevitch et al. (1992),
and others.

Inhomogeneities of the IGM will translate into CMB {\it fluctuations}
$\Delta T/T \sim y/\sqrt{N}$, $N$ being the effective number of blobs
per beam. Detailed maps of fluctuations expected in the framework of
self similar evolution of clusters of galaxies have been produced by
Markevitch et al. (1992).

Fluctuations $\delta T/T \sim 10^{-5}$ on scales $\theta \sim 2'$ may be
expected from correlated motions of plasma concentrated in young galaxies
or pre-galactic star clusters; anisotropies of similar or somewhat larger
amplitude but on smaller scales ($\theta \sim 3\Omega\,$arc sec) can be
produced by scattering of microwave photons by moving plasma in young
galaxies (Peebles 1990).

\subsection{Astrophysical backgrounds}

A particularly intense  far-IR/sub--mm background is predicted if
a large fraction of the hard XRB comes from starburst galaxies (Griffiths
\& Padovani 1990), owing to the relatively low efficiency of these
sources in producing high energy photons.
{\it Einstein Observatory} data indicate that
they generally have $(\nu I_\nu)_{\rm 2 keV}/(\nu I_\nu)_{60
\mu{\rm m}}  \lsim 10^{-4}$ (Fabbiano 1990). Thus models implying that
they make up a
large fraction of the hard XRB, might face problems with
the upper limits on the far-IR isotropic flux
as well as with energy constraints following from
estimates of the density of metals in galaxies (cf. De Zotti et al. 1991).

Things turn out to be more or less right
if the active starforming phases make up $\simeq 10\%$ of the XRB at 2 keV.
The contribution at lower energies might well be significantly higher
if these sources have the steep soft X-ray spectra
indicated by ROSAT data (Boller et al. 1992).

\medskip\noindent
A substantial far-IR/sub-mm background is, however, expected
also from galaxies directly detected in very deep optical
and near-IR surveys. The integrated flux corresponding to
direct counts in the wavelength range $\lambda =$3200--10000 {\AA} is
(Tyson 1990) $\nu I_\nu \simeq 10^{-9}\,\hbox{erg}\,\hbox{cm}^{-2}\,
\hbox{s}^{-1}\,\hbox{deg}^{-2}$
i.e. $\simeq 5\times 10^{-3}(\nu I_\nu)_{{\rm CMBpeak}}$.
The observed flattening of the counts suggests that the total contribution
of galaxies is not much larger than that.
The local far-IR luminosity density of galaxies is about \frac1/3
of the optical luminosity density (Saunders et al. 1990);
substantial cosmological evolution in the far-IR is suggested by
$60\,\mu$m IRAS counts (Hacking et al. 1987; Danese et al. 1987).
All that boils down to an expected far-IR/sub-mm background due to
directly observed galaxies
$(\nu I_\nu)_g \simeq \hbox{few}\times 10^{-3}(\nu I_\nu)_{{\rm CMBpeak}}$,
peaking at $\lambda \simeq 100(1+z_{\rm eff})\,\mu$m
(Franceschini et al. 1991).

\medskip\noindent
The estimated energy density of known AGNs,
$\epsilon_{\rm AGN} \approx 5\times 10^{-16}\,\hbox{erg}\,\hbox{cm}^{-3}$,
is $\approx 10^{-3} \epsilon_{\rm CMB}$ and
corresponds to a mass density of collapsed nuclei of
$(H_0/50)^2\Omega_{\rm AGN} \simeq 3\times 10^{-6}
(\kappa/30)\linebreak(\eta/0.1)^{-1}$,
where $\eta$ is the mass-energy conversion efficiency (Padovani et al. 1990).
A similar mass density of dust-enshrouded AGNs accreting with the normally
adopted efficiency $\eta \simeq 0.1$ would yield a far-IR background
potentially detectable by COBE. Already available data
on diffuse backgrounds
rule out the possibility that the dark matter consists of black holes
built up by accretion with such efficiency (see Bond et al. 1991).

\medskip\noindent
The possibility that
early structures, at $z \sim 5$--100, could have led to copious star
formation, producing both an intense background and dust capable of
reprocessing it, has been extensively discussed by Bond et al. (1991).
In this case, essentially all the energy produced by nuclear reactions
comes out at far-IR/sub-mm wavelengths. The peak wavelength depends
on the redshift and temperature distributions of the dust but, for
a relatively broad range of parameter values, occurs at $\lambda
\sim 600\,\mu$m.

\medskip\noindent
Measurements of the far-IR background spectrum would be informative on
the birth of galaxies, their chemical and photometric evolution,
the evolution of interstellar dust, the average density of metals in
the universe.
On the other hand, spectral measurements alone will not
identify the sources of the background; measurements of small scale
anisotropies would provide complementary information (Bond et al. 1991;
Peebles 1990).
For example, surface brightness fluctuations would be
suppressed if there is a large contribution from star clusters
not (yet) bound in galaxies; also, if the far-IR background is dominated
by thermal radiation from dust in galaxies, the autocorrelation function
of the far-IR intensity fluctuations directly reflects the galaxy correlation
function.

\section{The near-IR/optical background}

Measurements are difficult because of the intense
foreground emissions. In fact, the observational situation
is still unclear. The tightest upper limits
are already
only a factor of several above the expected contribution from galaxies
and about a factor of 10 above the integrated light from direct counts.
Claims of a substantially more intense
background in some bands are not confirmed by
more recent experiments (Noda et al. 1992).

The flattening of the QSO counts at $B\simeq 19.5$ suggests that
their contribution to the optical/near-IR background is small in comparison
to that of galaxies.

Predictions for a radiating highly ionized IGM at $T\sim 10^4\,$K
are at least two orders of magnitude below the emission from galaxies
(Paresce et al. 1980).

Measurements of the autocorrelation function of
optical and near-IR fluctuations have placed important
constraints on the background intensity.
A determination of the fluctuation spectrum would provide a test for the
presence of sources other than galaxies and measures at different angular
scales would probe the space distribution of sources (see Peebles 1990
and references therein).

\section{The UV background}

Its magnitude has been the subject of dispute for many years
(see Bowyer 1991; Henry 1991).

The first, and still one of the strongest pieces of evidence of an intense
diffuse ionizing flux
is the Gunn-Peterson effect. The lack of continuous absorption
shortward of the Ly$\alpha$ line in distant QSOs implies that the IGM
must be highly ionized up to $z\simeq 4.9$ (Schneider et al. 1991).

The ``proximity'' or ``inverse'' effect (decrease in the counted number
of intergalactic Ly$\alpha$-absorbing clouds in the vicinity of luminous,
UV-emitting quasars) gives an independent estimate of
the ionizing background intensity,
$J_{912} \sim 10^{-21\pm 0.5}\,\hbox{erg}\,\hbox{cm}^{-2}\,\hbox{s}^{-1}
\,\hbox{Hz}^{-1}\,\hbox{sr}^{-1}$
at the hydrogen Ly$\alpha$ edge (912 \AA), approximately independent
of redshift for $1.7 < z < 3.8$ (Lu et al. 1991).

Measurements of the autocorrelation function of intensity fluctuations
in the 1400--1900 {\AA} band (Martin \& Bowyer 1989) have allowed to
derive tight upper limits on the background flux.

{\it Young galaxies} could easily produce the UV flux up to 4 Ryd, if
enough radiation can escape from the galaxy (a galaxy undergoing a
burst of star formation is obviously expected to be gas-rich)
and a significant fraction of the metal abundance is produced
before $z\simeq 3$ (Miralda-Escud\'e \& Ostriker 1990 and references
therein).
Also Steidel \& Sargent (1989) have argued that to account for
the observed ionization state of heavy element absorption systems
a harder UV spectrum than produced by hot main sequence stars is required;
on the other hand, such ionization state may also be affected by
local sources of radiation or by collisional ionization
(Miralda-Escud\'e \& Ostriker 1990).

The integrated UV background from {\it observed QSOs} as a function of
redshift has been discussed by many authors (Madau, 1992
and references therein). The general conclusion
is that optically selected QSOs probably
 fail to emit sufficient ionizing flux
to account for the ionization level implied by the Gunn-Peterson test
and for the proximity effect, particularly at $z\gsim 2.5$ where a decline in
the comoving space density is suggested.

AGNs could nevertheless be the sources of the UV background.
The missing contribution may come either from quasars not seen because of
dust obscuration by intervening galaxies (Miralda-Escud\'e \& Ostriker
1990) or by a new class of AGNs such as the reflection dominated ones,
proposed to be the source of the hard XRB (Fabian et al. 1990).
The model discussed by Fabian et al. (1990)
predicts $J_{912} \sim 10^{-21}\,\hbox{erg}\,\hbox{cm}^{-2}\,\hbox{s}^{-1}
\,\hbox{Hz}^{-1}\,\hbox{sr}^{-1}$ at $z\simeq 2$--3.

{\it Protogalactic shock radiation\/} appears to be insufficient
by a large factor to account for the required photoionization (Shapiro \&
Giroux 1989).
{\it Decaying 'inos\/} as possible sources of the UV background are
discussed e.g. by Field \& Walker (1989).

\section{The X-ray background (XRB)}

\subsection{XRB constraints on evolution of the AGN population}

About 50\% of the 1--2 keV XRB has been resolved into discrete sources
in the deepest ROSAT field (Hasinger 1992), implying that X-ray surveys
are seeing directly almost the entire evolution history of X-ray sources
and in particular of AGNs. However, the situation is still somewhat
confusing mostly because data from different X-ray bands yield apparently
contradictory results (see Franceschini et al. 1992 for additional details
and references).

\begin{itemize}

\item HEAO-1 and {\it Ginga} fluctuation analyses
indicate a normalization of source counts
a factor of about 3 above that derived from the {\it Einstein Observatory}
EMSS, a result confirmed by direct {\it Ginga} counts.

\item The energy spectrum of fluctuations in the {\it Ginga} band (4--12 keV)
is consistent with the ``canonical'' AGN slope $\alpha \simeq 0.7$
(but beware of the effect of clusters!) and
substantial photoelectric absorption ($N_H \sim 10^{22}\,{\rm cm}^{-2}$).

\item On the other hand, soft X-ray selected AGNs rather show steep spectral
indices ($\alpha \simeq 1$--1.3) and no significant absorption
down to faint flux limits. Also, there is a distinct absence, at faint X-ray
fluxes, of the low luminosity AGNs which dominate the HEAO-1 A2 LLF
by Piccinotti et al. (1982).

\end{itemize}

Allowing for the presence of different components (a self absorbed
power law plus a soft excess plus a high energy bump) with a broad
distribution of spectral parameters is not enough to account for all the data
(the most critical being the mean spectral index of AGNs detected in deep
ROSAT surveys), as far as we assume: i) continous distributions and
ii) no spectral evolution.

A consistent picture obtains considering {\it two} AGN populations
(see Franceschini et al. 1992 for details).

\begin{itemize}

\item A {\it soft X-ray spectrum population}, contributing $\lsim 30\%$
of local hard X-ray selected AGNs (Piccinotti et al. 1982), of relatively
high luminosity, strongly evolving, as indicated by soft X-ray counts,
with X-ray spectra similar to those of optically selected quasars.

\item A {\it hard X-ray spectrum population}, contributing most of the
AGN in the Piccinotti survey, having a ``canonical'' spectral index,
strong self absorption ($N_H \sim 10^{22}\,{\rm cm}^{-2}$)
plus a high energy bump, relatively low mean luminosity. Since no more
than $\simeq 10$--15\% of EMSS and few ROSAT AGNs can belong to this class,
they should either be evolving very slowly (if at all) or be characterized
by a spectral evolution essentially counterbalancing,
in soft X-ray bands, their luminosity/density evolution. The latter possibility
is favoured by fluctuation analyses;
evolution is required if these sources have to
account for the XRB intensity above 3 keV.

\end{itemize}

{\it Soft}-spectrum AGNs should give a minor contribution to the
XRB above 3 keV (the model discussed by Franceschini et al. 1992 yields
23\% at 3 keV) and might also not saturate the XRB intensity in soft
bands. In fact, at least in the framework of luminosity evolution models,
the decline of their local luminosity function at low luminosities,
indicated by {\it Einstein Observatory} Medium Sensitivity Survey data,
entails a correspondingly fast convergence of the counts and an
integrated flux at 1 keV $\simeq 30$--50\% of the XRB intensity estimated
from ROSAT data (cf. Franceschini et al. 1992).
Additional contributions from {\it soft} X-ray sources
(IGM in groups and clusters, ASF galaxies...) may be required.

{\it Hard}-spectrum AGNs are essentially invisible below a few keV.
They could account for the XRB at higher energies if both their
luminosity and their absorbing columns evolve.

A third AGN population characterized by extreme absorbing columns
($N_H \sim 10^{24}\hbox{--}10^{25}\,{\rm cm}^{-2}$; Setti \& Woltjer 1989)
or by a reflection spectrum (Fabian et al. 1990)
could be the dominant contributor to the XRB above 3 keV. Optically
selected Seyfert 2 may have the necessary properties.
This population, however, is not represented in the Piccinotti LLF
and apparently cannot give a large contribution to fluctuations
measured by {\it Ginga} [Tanaka (1992) quotes an upper limit
of $\sim 20\%$ to this contribution down to
$S_{\rm 2-10 keV} = 2\times 10^{-13}\,\hbox{erg}\,\hbox{cm}^{-2}\,
\hbox{s}^{-1}$], implying
a very low local volume emissivity, so that extreme evolution is
needed to produce a significant fraction of the XRB.

\subsection{XRB constraints on large scale structure }

If the minimum angle, $\theta_{min}(r_0)$, subtended by the
maximum scale of significant clustering is larger than the angular scale
of observations, the autocorrelation function $W(\theta)$ scales as
(De Zotti et al. 1990)
$W(\theta) \propto \theta^{1-\gamma} r_0^\gamma f^2$,
where $r_0$ is the ``clustering radius'' at the present time ($\xi(r) =
(r/r_0)^{-\gamma}$),
$f$ is the fraction of the residual background (after subtraction
of the resolved source contribution) produced by the considered population.

If, on the other hand, $\theta_{min}(r_0) \ll \theta$, most of the
contribution to the observed ACF comes from local sources, which
provide only a minor fraction of the XRB; the above equation
still holds but $f$ has to be replaced by the fraction of the XRB volume
emissivity made up by the considered sources: $j_{\rm sources}/j_{\rm XRB}$.
Particularly tight constraints are obtained for AGNs, whose local
volume emissivity is $\simeq 20\%j_{\rm XRB}$
(Barcons \& Fabian 1989; Mart{\'\i}n-Mirones et al. 1991;
Carrera \& Barcons 1992).

Several analyses of clustering properties of optically selected
quasars consistently indicate a 2-point correlation function consistent
with the $-1.8$ power law form derived for galaxies with a scale
length $r_0 = (14 \pm 3)(50/H_0)\,$Mpc at $z \simeq 1.4$
(Andreani et al. 1991, and references therein).
The availability of large quasar samples has recently made possible
to investigate the cosmological evolution of their clustering. Such
evolution is usually modelled as
$\xi(r,z) = (r/r_0)^{-\gamma}(1+z)^{-(3+\epsilon)}$,
where $\epsilon = 0$ corresponds to ``stable clustering'',
$\epsilon = \gamma -3$  or $\epsilon = -3$ to a clustering radius constant in
comoving or in physical coordinates, respectively, $\epsilon = \gamma -1$
to linear growth for $q_0=0.5$.
Recent studies (Andreani \& Cristiani 1992) support
the ``comoving'' model ($\epsilon = \gamma -3 \simeq - 1.2$), although
stable clustering, favoured by some earlier analyses, cannot yet be ruled out.

But sources with $r_0 > 10\,$Mpc and $\epsilon = -1.2$ would produce
an X-ray ACF exceeding the small scale ROSAT limits (Hasinger 1992), if
their contribution to the XRB exceeds $\simeq 30\%$. Sources
with $r_0 \geq 10\,$Mpc can account for $\geq 50\%$ of the XRB only if
$\epsilon \geq 0$.
Similar constraints apply to Active Star-forming Galaxies.
If their clustering radius is equal to that of normal galaxies
($r_0 \simeq 10\,$Mpc), their contribution to the soft XRB cannot exceed
$\simeq 30\%$ if $\epsilon = \leq -1.2$ and is $\lsim 45\%$ if $\epsilon = 0$.

Mushotzky \& Jahoda (1992) report the detection (99\% confidence)
of a positive ACF at scales $\sim 6^\circ$--$20^\circ$ with $W(\theta)
\sim 3\times 10^{-5}$. Based on the LLFs by Piccinotti et al. (1982),
Danese et al. (1992) find that rich clusters with $r_0 = 50\,$Mpc
yield $W(6^\circ) \simeq 10^{-5}$, while AGNs with $r_0 = 20\,$Mpc
give $W(6^\circ) \simeq 3\times 10^{-5}$. Any class of sources
distributed like normal galaxies ($r_0\simeq 10\,$Mpc)
must have a local volume emissivity $j_{\rm gal} \lsim 0.4 j_{\rm XRB}$.

\medskip\noindent
{\it Acknowledgements.} GDZ wishes to thank the organizers
for having allowed him to attend this very
successful workshop and the Pontificial Academy of Sciences for the
very warm hospitality. Work supported in part by ASI and CNR.

\medskip
\centerline{\bf References}

\def\ref{\noindent\hangindent=20pt\hangafter=1}

\ref
Andreani, P., Cristiani, S., \& La Franca, F. 1991, MNRAS, 253, 527

\ref
Andreani, P., \& Cristiani, S. 1992, ApJ, 398, L13

\ref
Barcons, X., \& Fabian, A.C. 1989, MNRAS, 237, 119

\ref
Bignami, G.F. 1992, Nature, 360, 416

\ref
Boller, Th., Meurs, E.J.A., Brinkmann, W., Fink, H., Zimmermann, U.,
\& Adorf, H.-M. 1992, A\& A, 261, 57

\ref
Bond, J.R., Carr, B.J., \& Hogan, C.J. 1991, ApJ, 367, 420

\ref
Bowyer, S. 1991, ARAA, 29, 59

\ref
Carrera, F.J., \& Barcons, X. 1992, MNRAS, 257, 507

\ref
Cavaliere, A., Menci, N., \& Setti, G. 1991, A\& A, 245, 59

\ref
Cen, R., \& Ostriker, J. 1992, ApJ, 393, 22

\ref
Danese, L., De Zotti, G., Franceschini, A., \& Toffolatti, L. 1987, ApJ,
318, L15

\ref
Danese, L., De Zotti, G., \& Andreani, P. 1992, in The X-ray Background,
Cambridge Univ. Press, p. 61

\ref
De Zotti, G., Persic, M., Franceschini, A., Danese, L., Palumbo, G.G.C.,
Boldt, E.A., \& Marshall, F.E. 1990, ApJ, 351, 22

\ref
De Zotti, G., Mart{\`\i}n-Mirones, J.M., Franceschini, A., \&
Danese, L. 1991, in Proc. XI Moriond Astrophysics Meeting ``The Early
Observable Universe from Diffuse Backgrounds'', p. 31

\ref
Fabbiano, G. 1990, ARAA, 27, 87

\ref
Fabian, A.C., George, I.M., Miyoshi, S., \& Rees, M. 1990, MNRAS, 242, 14P

\ref
Field, G.B., \& Perrenod, S.C. 1977, ApJ, 215, 717

\ref
Field, G.B., \& Walker, T.P. 1989, Phys. Rev. Lett. 63, 117

\ref
Franceschini, A., Mart{\'\i}n-Mirones, J.M., Danese, L., \& De~Zotti, G.
1992, MNRAS, submitted

\ref
Franceschini, A., Toffolatti, L., Mazzei, P., Danese, L., \& De~Zotti, G.
1991, A\& A Suppl. 89, 285

\ref
Griffiths, R.E., \& Padovani, P. 1990, ApJ, 360, 483

\ref
Hacking, P.B., Condon, J.J., \& Houck, J.R. 1987, ApJ, 316, L15

\ref
Hasinger, G. 1992, in X-ray Emission from Active Galactic Nuclei and the
Cosmic X-ray Background, MPE Report 235, p. 321

\ref
Henry, R.C. 1991, ARAA, 29, 89

\ref
Longair, M.S., \& Sunyaev, R.A. 1971, Usp. Fiz. Nauk, 105, 41 (Sov. Phys. Usp.
14, 569, 1972)

\ref
Lu, L., Wolfe, A.M., \& Turnshek, D.A. 1991, ApJ, 367, 19

\ref
Madau, P. 1992, ApJ, 389, L1

\ref
Markevitch, M., Blumenthal, G.R., Forman, W., Jones, C., \& Sunyaev, R.A.
1992, ApJ, 395, 326

\ref
Martin, C., \& Bowyer, S. 1989, ApJ, 338, 677

\ref
Mart{\'\i}n-Mirones, J.M., De~Zotti, G., Boldt, E.A., Marshall, F.E.,
Danese, L., Franceschini, A., \& Persic, M. 1991, ApJ, 379, 507

\ref
Miralda-Escud\'e, J., \& Ostriker, J.P. 1990, ApJ, 350, 1

\ref
Mushotzky, R., \& Jahoda, K. 1992, in The X-ray Background,
Cambridge Univ. Press, p. 80

\ref
Noda, M., et al. 1992, ApJ, 391, 456

\ref
Padovani, P., Burg, R., \& Edelson, R.A. 1990, ApJ, 353, 438

\ref
Paresce, F., McKee, C., \& Bowyer, S. 1980, ApJ, 240, 387

\ref
Peebles, P.J.E. 1990, in Proc. IAU Symp. No. 139 ``The Galactic and
Extragalactic Background Radiation'', p. 295

\ref
Piccinotti, G., Mushotzky, R.F., Boldt, E.A., Holt, S.S., Marshall, F.E.,
Serlemitsos, P.J., \& Shafer, R.A. 1982, ApJ, 253, 485

\ref
Saunders, W., Rowan-Robinson, M., Lawrence, A., Efstathiou, G., Kaiser, N.,
Ellis, R.S., \& Frenk, C.S. 1990, MNRAS, 242, 318

\ref
Schneider, D.P., Schmidt, M., \& Gunn, J.E. 1991, ApJ, 306, 411

\ref
Setti, G., \& Woltjer, L. 1989, A\& A, 224, L21

\ref
Shapiro, P.R., \& Giroux, M.L. 1989, in The Epoch of Galaxy Formation,
Kluwer, p. 153

\ref
Songaila, A., Cowie, L.L., \& Lillie, S.J. 1990, ApJ, 348, 371

\ref
Steidel, C.C., \& Sargent, W.L.W. 1989, ApJ, 343, L33

\ref
Tanaka, Y. 1992, in X-ray Emission from Active Galactic Nuclei and the
Cosmic X-ray Background, MPE Report 235, p. 303

\ref
Tyson, J.A.  1990, in Proc. IAU Symp. No. 139 ``The Galactic and
Extragalactic Background Radiation'', p. 245

\end{document}